\documentclass[fleqn,twoside]{article}
\usepackage[headings]{espcrc2}

\readRCS
$Id: espcrc2.tex,v 1.2 2004/02/24 11:22:11 spepping Exp $
\usepackage{graphicx}
\usepackage[figuresright]{rotating}

\newcommand{\AmS}{{\protect\the\textfont2
  A\kern-.1667em\lower.5ex\hbox{M}\kern-.125emS}}
\newcommand{\Veff}{{\cal V}}	
\newcommand{\Aeff}{{\cal A}}
\newcommand{\lagr}{{\cal L}}
\newcommand{\calo}{{\cal O}}

\title{\boldmath{$K^L_{\mu3}$} decay: A first evidence of Right-Handed Quark Currents ?}

\author{E.Passemar\thanks{In collaboration with V.Bernard, M.Oertel,~J.Stern~\cite{Bernard:2006gy}.}
\address[MCSD]{Groupe de Physique Th\'{e}orique, IPN,
           Universit\'{e} de Paris Sud-XI, F-91406 Orsay, France\\
email: passemar@ipno.in2p3.fr}}
       
\begin{document}

\begin{abstract}
\noindent The experimental results published by KTeV and the preliminary results from NA48 
concerning the slope of the $K\pi$ scalar form factor suggest a 
significant discrepancy with the prediction of the Callan-Treiman 
low energy theorem once interpreted within the Standard Model. 
In this talk, we will show how this discrepancy could be explained as a first 
evidence of the direct coupling of right-handed quarks to 
W as suggested by certain type of effective electroweak theories.
\end{abstract}

\maketitle
\vspace*{-1.cm}
\section{Slope of the scalar $K\pi$ form factor.}
\begin{figure}[h]
\begin{center}
\vspace{-0.9cm}
\includegraphics*[width=5.5cm,clip]{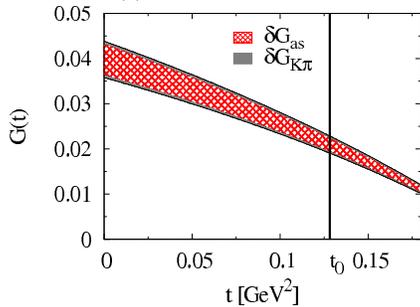}
\vspace{-1.1cm}
\caption{\small{G(t) with the uncertainties $\delta \mathrm{G}_{as}$ 
and $\delta \mathrm{G}_{K \pi}$ added in quadrature.}}
\label{figureg}
\end{center}
\vspace{-1.2cm}
\end{figure}

\vspace{-0.2cm}
\noindent The hadronic matrix element associated with $K^0_{\mu3}$ decay is
given by
\vspace{-0.2cm}
\begin{equation}
\begin{array}{lr}
\langle \pi^-(p') | \bar{s}\gamma_{\mu}u | K^0(p)\rangle = \\
(p'+p)_\mu\  f^{K^0\pi^-}_+ (t) + (p-p')_\mu\  f_-^{K^0\pi^-}(t)~,         
\label{hadronic element}
\end{array}
\end{equation}
\vspace{-0.3cm}
where $t=(p'-p)^2$. The vector form factor $f^{K^0\pi^-}_+ (t)$ represents
the P-wave projection of the crossed channel matrix element
$\langle 0 |\bar{s}\gamma_{\mu}u | K\pi \rangle$, whereas                
the S-wave projection is described by the scalar form factor
\vspace{-0.25cm}
\begin{equation}
f^{K^0\pi^-}_S (t) = f^{K^0\pi^-}_+ (t) + \frac{t}{m^2_K - m^2_\pi} f^{K^0\pi^-}_-(t)~.
\label{defffactor}
\end{equation}
\vspace{-0.2cm}
In the sequel we consider the normalized scalar form factor
\vspace{-0.6cm}    
\begin{equation}
~~~~~~~~~~~~~~~~f(t)=\frac{f^{K^0\pi^-}_S(t)}{f^{K^0\pi^-}_+(0)}\ \ ,\ \ f(0)= 1~.
\label{normff}
\end{equation}
\vspace{-0.28cm}
The experimental measurements usually concern the slope $\lambda_0$ and/or 
the curvature $\lambda_0'$ of the form factor considering a Taylor expansion
\vspace{-0.25cm}
\begin{equation}
f(t) = 1 + \lambda_0 \frac{t}{m_{\pi}^2} + \frac{1}{2} \lambda_0'
(\frac{t}{m_{\pi}^2})^2  + \ldots~.
\label{taylor}
\end{equation}
\vspace{-0.2cm}
A linear fit leads to the following values of the slope
\vspace{-0.65cm}
\begin{equation}
~~~~~~~~\lambda_{0}^{lin}=0.01372 \pm 0.00131~~~[\mathrm{KTeV}~\cite{Alexopoulos:2004sy}]~,
\label{slopeKTeV}
\end{equation}
\vspace{-0.25cm} 
\vspace{-0.29cm}
\begin{equation}
~~~~~~~~\lambda_{0}^{lin} = 0.0120 \pm 0.0017~~~[\mathrm{NA48}~\cite{Winhart:2005ig}]~.
\label{slopeNA48}
\end{equation}
\vspace{-0.30cm}
These have to be compared with the 
theoretical prediction of the Standard Model (SM). 
This can be done by matching the dispersive representation of the scalar form factor
with the Callan-Treiman (CT) low energy 
theorem~\cite{Dashen:1969bh} which predicts the value of $f(t)$ 
at the Callan-Treiman point $t=\Delta_{K \pi}=m_{K^{0}}^2-m_{\pi^+}^2$ in the 
$\mathrm{SU}(2)\times \mathrm{SU}(2)$
chiral limit. One has 
\vspace{-0.8cm}
\begin{equation}
\mathrm{C}\equiv f(\Delta_{K\pi})=\frac{F_{K^+}}{F_{\pi^+}}\frac{1}{f_{+}^{K^0\pi^-}(0)}+  
\Delta_{CT}
\label{C}
\end{equation} 
\vspace{-0.3cm}
where the CT correction, $\Delta_{CT} \sim \calo\Big{(}\frac{m_{u,d}}{4 \pi F_\pi}\Big{)}$,
is not enhanced by chiral logarithms or by small denominators arising 
from the $\pi^0$-$\eta$
mixing in the final state\footnote{This is not the case for the charged K decay mode where an extra contribution
due to the $\pi^0$-$\eta$ 
mixing in the final state is involved.}.
This correction has been estimated within Chiral Perturbation Theory (ChPT) at next to
leading order (NLO)
in the isospin limit 
\vspace{-0.15cm}
\begin{equation} 
\mathrm{\cite{Gasser:1984ux}~with~the~result:}~\Delta^{NLO}_{CT}=-3.5~10^{-3}~.
\label{Delta_CT}
\end{equation}
\vspace{-0.15cm}
Assuming the SM couplings, the experimental results for the branching ratios (BR) 
Br$\left({K^+_{l2}(\gamma)}/{\pi^+_{l2}(\gamma)}\right)$~\cite{Jamin:2006tj},
for $|f_+^{K^0\pi^-}(0) V^{us}|$~\cite{Alexopoulos:2004sw} 
and for $V^{ud}$~\cite{Marciano:2005ec} allow to write
\vspace{-0.15cm}
\begin{equation}
\begin{array}{rl}
\mathrm{C_{SM}} &=\Bigl{|}\frac{F_{K^+} V^{us}}{F_{\pi^+} V^{ud}}\Bigr{|}
\frac{1}{|f_+^{K^0\pi^-}(0) V^{us}|}|V^{ud}|+ \Delta_{CT}\nonumber\\
&= B_{exp} + \Delta_{CT}~, 
\label{CSM}
\end{array}
\end{equation}
\vspace{-0.15cm}
with $B_{exp}=1.2440 \pm 0.0039$~.
In the following, the relevant quantity will be 
\vspace{-0.2cm}
\begin{equation}
\mathrm{ln C_{SM}} = 0.2183 \pm 0.0031 + \Delta_{CT}/B_{exp}~.
\label{lnCSM}
\end{equation}
\vspace{-0.2cm}
Since we know the
value of $f(t)$ at two points at low energy: at $t=0$, Eq.~(\ref{normff}),
and at $t=\Delta_{K\pi}$, Eq.~(\ref C), one can write
a dispersion relation with two subtractions for ln($f(t)$). Assuming that 
$f(t)$ has no zero, one obtains
\vspace{-0.20cm}
\begin{equation}
f(t)=\exp\Bigl{[}\frac{t}{\Delta_{K\pi}}(\mathrm{lnC}-\mathrm{G(t)})\Bigr{]}~,~~~~\mathrm{where}
\label{Dispf}
\end{equation}
\vspace{-0.4cm}
\begin{eqnarray}
\mathrm{G(t)}&=&\frac{\Delta_{K\pi}(\Delta_{K\pi}-t)}{\pi}\nonumber\\
& \times &\int_{t_{K\pi}}^{\infty}
\frac{ds}{s}
\frac{\phi(s)}
{(s-\Delta_{K\pi})(s-t-i\epsilon)}~,
\label{G}
\end{eqnarray}
$t_{K\pi}$ is the threshold of $\pi K$ scattering and $\phi(t)$ is the
phase of $f(t)$.
According to Brodsky-Lepage, 
$f(t)$ vanishes as 
$\calo(1/t)$ for large t~\cite{Lepage:1979zb}, implying that 
$\phi(t) \stackrel{t\mapsto\infty}{\longmapsto } \pi$. $\mathrm{G(t)}$ can be decomposed into two parts:\\
$\mathrm{G}(t)=\mathrm{G}_{K\pi}(\Lambda,t) + \mathrm{G}_{as}(\Lambda,t) \pm \delta \mathrm{G}(t)$.
The first part, $\mathrm{G}_{K\pi}(\Lambda,t)$,  corresponds to the integration region $t_{K\pi} \leq s \leq \Lambda$
where the $\pi K$ S-wave is still observed to be elastic 
($\Lambda \simeq 2.77~\mathrm{GeV^2}$~\cite{Aston:1987ir},~\cite{Buettiker:2003pp}). 
In this region, $\phi(t)$ equals to the $I=1/2$ S-wave scattering phase shift 
according to Watson's theorem.
The scattering phase has been inferred from
experimental data~\cite{Aston:1987ir} solving the Roy-Steiner equations~\cite{Buettiker:2003pp}.
The second part, $\mathrm{G}_{as}(\Lambda,t)$, is
the asymptotic contribution to the integral, Eq.~(\ref{G}),
for $s>\Lambda$. There, we replace $\phi(t)$ by its asym-ptotic value $\pi$.
We include the possible devia-tion from this asymptotic estimate into the uncertainty.
Thanks to the two subtractions,
the integral in Eq.~(\ref G) converges very 
rapidly and $\mathrm{G}_{K\pi}(\Lambda,t)$ dominates. $\delta \mathrm{G}(t)$ contains the two sources of uncertainties arising from
the two parts of $\mathrm{G}(t)$ as discussed in details in~\cite{Bernard:2006gy}.
The resul-ting function $\mathrm{G}(t)$ is shown in Fig.~\ref{figureg}.
\\Using the exact parametrization, Eq.~(\ref{Dispf}), the linear slope and the curvature are given by
\vspace{-0.25cm}
\begin{equation}
\lambda_0 = m_{\pi}^2~f'(0)= \frac{m_{\pi}^2}{\Delta_{K \pi}}(\mathrm {ln C} - \mathrm{G(0)})~,
\label{slope}
\end{equation}
\vspace{-0.55cm}
\begin{equation}
\begin{array}{rl}
\lambda_0' &= m_{\pi}^4~f''(0)=\lambda_0^2 - 2 \frac{m_{\pi}^4}{\Delta_{K\pi}} \mathrm{G'(0)} \nonumber\\
&= \lambda_0^2  + (4.16 \pm 0.50)\times 10^{-4}~.
\label{curvature}
\end{array}
\end{equation}
\vspace{-0.35cm}
Taking the value of $\mathrm{lnC_{SM}}$, Eq.~(\ref{lnCSM}), 
we obtain\footnote{This, in principle, 
increases the precision of the NLO ChPT result, $\lambda_0=0.017 \pm 0.004$~\cite{Gasser:1984ux}.}:
\vspace{-0.65cm}
\begin{equation}
\lambda_0 = 0.01524 \pm 0.00044 + 0.0686~\Delta_{CT}~,
\label{lambda}
\end{equation}
\vspace{-0.25cm}
to be
compared with the experimental result of KTeV, Eq.~(\ref{slopeKTeV}), and 
the preliminary one of NA48, Eq.~(\ref{slopeNA48}).
With the estimate of $\Delta_{CT}^{NLO}$, Eq.~(\ref{Delta_CT}),
the KTeV result is still compatible with the theoreti-cal 
prediction, whereas the NA48 result requires $\Delta_{CT} \leq  -2.2~10^{-2}$, i.e. 
at least six times larger in absolute value than the estimate of Eq.~(\ref{Delta_CT}).
Moreover these measurements
do not take into account the effect of the \textbf{positive curvature} $\lambda_0'$, Eq.~(\ref{curvature}),
in a proper way. For this reason, they should be actually interpreted as representing an \textbf{upper bound}
for $\lambda_0=m_{\pi}^2~f'(0)$, Eq.~(\ref{lambda}), since the curvature is necessary positive. This, concerning 
the NA48 result at least, accentuates the discrepancy between the experimental measurements
and the SM prediction of $\lambda_0$. 
The actual value of $\lambda_0$ should be confirmed 
by the direct measurement of lnC using the exact dispersive parametrization, Eq.~(\ref{Dispf}). This parametrization
 is very powerful since one parameter, lnC, allows a measurement
of both the slope and the curvature of $f(t)$.
In this way, one avoids the problem of the strong correlations as shown by Eq.~(\ref{curvature}) 
that appears in the extraction of the slope and the curvature
using the quadratic parametrization, Eq.~(\ref{taylor}).
\vspace{-0.6cm}
\section{A first evidence of right-handed quark currents (RHCs)~?}
\begin{figure}[h!]
\begin{center}
\includegraphics*[width=6.5cm,clip]{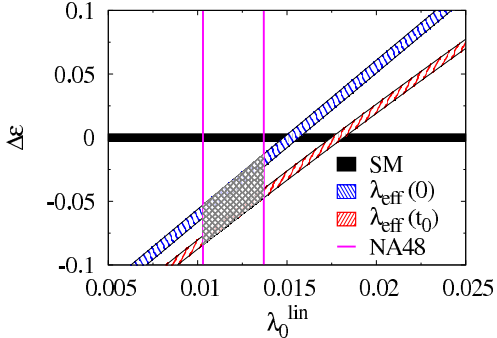}
\vspace{-1.1cm}
\caption{\small {Impact of NA48 data on RHCs. Horizontal line: SM, $\Delta \epsilon = \Delta_{CT}^{NLO}/B_{exp} = \pm~0.0028$,
vertical lines: NA48 measurements of $\lambda_0$. Top stripped curve: $\lambda_{0}^{lin}=\lambda_{eff}(0)$  and bottom
stripped curve: $\lambda_{0}^{lin}=\lambda_{eff}(t_0)$
with uncertainties from BR and from G(t) added in quadrature.}}    
\label{RHCs}
\end{center}
\vspace{-1.3cm}
\end{figure}

\vspace{-0.2cm}
\noindent 
We now point out how a possible discrepancy 
between the SM prediction of the linear slope $\lambda_0$ and its 
measurements could be interpreted as a mani-festation of physics beyond the SM.
We refer 
to the framework of the "low energy effective theo-ry" (LEET)
developed in \cite{Hirn:2005fr}. It is constructed by ordering 
all the vertices invariant under a sui-table symmetry group
according to their infrared (chiral) dimension $d$:
$\lagr_{eff} =  \Sigma_{d\geq 2} \lagr_{d}$, with the operators that behave as
$\lagr_{d}=\calo(p^d)$ in the low energy (LE) limit $p\ll \Lambda \sim  3~\mathrm{TeV}$. 
The LEET is not renormalized and unitarized in the usual sense, 
but order by order in the LE expansion.
In the LEET, as in other extensions of the SM, 
the heavy states beyond the SM present at high energy $(E>\Lambda)$ decouple. 
However the higher local symmetries
originally associated to them
that contain the SM gauge group as a sub-group do not decouple at LE: they survive and become 
non li-nearly realized restricting the interaction vertices of $\lagr_{eff}$.
This higher symmetry, $S_{nat}$, can be inferred~\cite{Hirn:2005fr}
from the SM itself: $S_{nat}$ is required to select at leading order (LO) ($\calo (p^2)$)
the higgs-less vertices of the SM and nothing else.
The minimal symmetry group that satisfies this condition is\footnote{
The discrete symmetry $Z_2$ ($\nu_R \rightarrow -\nu_R $) 
forbids the Dirac
masses of neutrinos 
and at the same time 
the leptonic charged RHCs. 
Consequently the stringent 
constraints, which come from polarization measurements 
in $\mu$, $\tau$ and $\beta$ decays
and 
occur in left-right symmetric models, are automatically satisfied.} 
$S_{nat}=[SU(2) \times SU(2)]^2 \times U(1)_{B-L} \times Z_2$~.
The reduction of this higher symmetry $S_{nat}$ to $SU(2)_L \times U(1)_{Y}$ 
is done via spurions~\cite{Hirn:2005fr}. Higher terms in $\lagr_{eff}$ are 
suppressed according to their infrared dimension $d$ and the number of spurions
that is needed to restore the invariance under $S_{nat}$. At LO ($\calo(p^2)$), we
recover the SM couplings without a physical scalar with fermions masses generated 
by spurions. 
The first and the most important effects of new physics appear at NLO,
before the loops and oblique corrections which only arise at NNLO.
At NLO, there are only two operators
instead of the 80 operators of mass dimension $\mathrm{D}=6$ characteristic 
of the usual decoupling scenario. 
These two operators modify the couplings of fermions to W and Z.
The charged current (CC) lagrangian becomes
\vspace{0.1cm}
$
\lagr_{CC}=\tilde{g}[l_\mu +\frac{1}{2}\bar{\mathrm{U}}( \Veff_{eff} \gamma_{\mu}  +
\Aeff_{eff}\gamma_{\mu} \gamma_5 ) \mathrm{D}] W^\mu+hc
$
\noindent where
$\mathrm{U}=
\left(\begin{array}{c}
u\\ c \\t 
\end{array}\right)$,~
$\mathrm{D}=
\left(\begin{array}{c}
d\\ s \\b 
\end{array}\right)$,\\
and $\Veff_{eff}$,~$\Aeff_{eff}$ 
are complex $3 \times 3$ effective      
coupling matrices. In the SM,
$\Veff_{eff} = - \Aeff_{eff} = V_{CKM}$~,
where $V_{CKM}$ is the unitary flavour mi-xing matrix,
whereas at NLO right-handed 
quark currents (RHCs) are present. Indeed,
\vspace{-0.2cm}
\begin{equation}
\begin{array}{lr}
\Veff_{eff}^{ij}~=&(1+\delta) V_L^{ij}+\epsilon V_R^{ij}+\mathrm{NNLO}~,\nonumber\\
\vspace{-0.1cm}
\Aeff_{eff}^{ij}~=&-(1+\delta) V_L^{ij}+\epsilon V_R^{ij}+\mathrm{NNLO}~,
\label{effective couplings}
\end{array}
\end{equation}
\vspace{-0.1cm}
with $V_L$ and $V_R$ two unitary flavor mixing matrices coming 
from the diagonalization of the mass matrix of U and D
quarks; $\delta$ and $\epsilon$ are small parameters originating from spurions 
which have been estimated~\cite{Hirn:2005fr}
of the order of one percent. $l_{\mu}$ in $\lagr_{CC}$ stands for the usual V-A leptonic current since
the discrete symmetry $Z_2$ forbids leptonic charged RHCs.
These new couplings,~Eq.~(\ref{effective couplings}), affect 
the reexpression of Eq.~(\ref{C}) in terms of measurable
BR leading to
$\mathrm{C} = B_{exp}~r +\Delta_{CT}$~,
where $B_{exp}$ has the same value as the one defined in Eq.~(\ref{CSM}). 
However in the presence of RHCs, it reads\\
$B_{exp}=\Bigl{|}\frac{F_{K^+} \Aeff_{eff}^{us}}{F_{\pi^+} \Aeff_{eff}^{ud}}\Bigr{|}
\frac{1}{|f_+^{K^0\pi^-}(0) \Veff_{eff}^{us}|}|\Veff_{eff}^{ud}|$ and an
additional factor $r$ appears. It is given in terms of 
RHCs effective couplings
\vspace{-0.2cm}
\begin{equation}
r = \Bigl{|}\frac{\Aeff_{eff}^{ud} \Veff_{eff}^{us}}{\Veff_{eff}^{ud}
\Aeff_{eff}^{us}}\Bigr{|} 
= 1 + 2 (\epsilon_{S}-\epsilon_{NS})+ \calo(\epsilon^2)~,
\label{r}
\end{equation}
\vspace{-0.2cm}
where
\vspace{-0.5cm}
\begin{equation}
\epsilon_{NS}= \epsilon\ \mathrm{Re} \Bigl{(}\frac{V_R^{ud}}{V_L^{ud}}\Bigr{)},\ \
\epsilon_{S} = \epsilon\ \mathrm{Re} \Bigl{(}\frac{V_R^{us}}{V_L^{us}}\Bigr{)}
\label{epsilon}
\end{equation}
\vspace{-0.2cm}
represent the strengths of $\bar ud$ and $\bar us$ RHCs, respectively. 
Hence Eq.~(\ref{lnCSM}) 
can be rewritten as
\vspace{-0.65cm}
\begin{equation}
\mathrm{ln C} = 0.2183 \pm 0.0031 + \Delta \epsilon~,
\label{lnC}
\end{equation}
\vspace{-0.25cm}
with $\Delta \epsilon = \Delta_{CT}/B_{exp} + 2 (\epsilon_S - \epsilon_{NS}) + \calo(\epsilon^2)$,
\\a combination of the RHCs couplings,\\ 
$\Delta \epsilon_0 = 2 (\epsilon_S - \epsilon_{NS})$, 
and the CT correction, $\Delta_{CT}/B_{exp}$.
As mentioned before, the
experimental measurements published so far only give an upper bound for $\lambda_0$
and hence for lnC and for $\Delta \epsilon$.
Comparing the experimental results of $\lambda_0$, Eq.~(\ref{slopeKTeV}) and 
Eq.~(\ref{slopeNA48}), with Eq.~(\ref{slope}) and using
Eq.~(\ref{lnC}), we obtain an upper bound estimate for $\Delta \epsilon$\footnote{The resulting
uncertainty is the quadratic sum of the uncertainties on $\lambda_0^{lin}$, on $B_{exp}$ and 
on G(t).}: 
\vspace{-0.25cm}
\begin{equation}
\Delta \epsilon_{max}=-0.0178 \pm 0.0161~~~[\mathrm{KTeV}]~,
\label{depsKTeV}
\end{equation}
\vspace{-0.5cm}
\begin{equation}
\Delta \epsilon_{max} =-0.0379 \pm 0.0205~~~[\mathrm{NA48}]~.
\label{depsNA48}
\end{equation}
\vspace{-0.35cm} 
Hence the result coming from the KTeV measurement, Eq.~(\ref{depsKTeV}),
can still be interpreted, to a certain extend,
within the SM with $\Delta \epsilon = \Delta_{CT}^{NLO}/B_{exp}$.
The result coming from the measurement of NA48, Eq.~(\ref{depsNA48}), 
seems more difficult to interpret as a pure CT correction, since that would require
$|\Delta_{CT}| \geq 6~|\Delta_{CT}^{NLO}|$. 
A non zero value of RHCs, $\Delta \epsilon_0$,
can provide an explanation.
Using the published experimental measurements based on the linear parametrization of $f(t)$, 
we can visualize the effect of RHCs. 
For this purpose, we define the effective slope
$\lambda_{eff}(t)=\frac{m_{\pi}^2}{t}\Big{[}f(t)-1\Big{]}$.
For each fixed t, $\lambda_{eff}(t)$ is a function of lnC Eq.~(\ref{Dispf}) or of $\Delta \epsilon$.
For the extreme cases $t = 0$ and $t = t_0=(m_{K^0}-m_{\pi^+})^2$,
these two curves are displayed in Fig.~\ref{RHCs} together with the range of $\lambda_0$ given by the NA48 measurement.
Since $\lambda_{eff}(t)$ is increasing with $t$, the measured value $\lambda_0^{lin}$ is between 
$\lambda_{eff}(0)= m_{\pi}^2~f'(0)$ and $\lambda_{eff}(t_0)$. Consequently the true value of $\Delta \epsilon$ 
should be somewhere in the gray stripped
region shown in Fig.~\ref{RHCs}
suggesting $\Delta \epsilon = - 0.05 \pm 0.03$. 
A similar figure in the case of KTeV can be found in~\cite{Bernard:2006gy} and leads to $\Delta \epsilon = - 0.03 \pm 0.03$.
One should wonder whether such a "large" effect of RHCs can be generated by genuine spurions parameters 
$\delta$ and $\epsilon$ of the size of one percent~\cite{Hirn:2005fr}.  
Since the left-handed mixing matrix is very close (equal at LO) to the CKM matrix,
we can neglect the $V_L^{ub}$ element. 
Hence $V_{L}^{ud}$ is of order one and $V_{L}^{us}\sim 0.22$. 
As the unitarity of $V_R$ forces $|V_R^{ud}|\leq 1$, $|\epsilon_{NS}|$ can hardly exceed $\epsilon$ and
therefore $|\epsilon_{NS}| \sim \epsilon \sim 1\%$.
On the other hand, because $V _L ^{us}$ is suppressed, $\epsilon _S$ is enhanced 
unless $V _R ^{us}$ is suppressed too. If the hierarchy in the 
right-handed sector is inverted, then $|\epsilon_S|$ can reach a value of order
$4.5~\epsilon$ and thus $|\Delta \epsilon_0|$ could be as large as nine percent.
\vspace{-0.2cm}
\section{Conclusion.}
\vspace{-0.2cm}
\noindent In this talk, we have pointed out a 
possible discrepancy between the SM and 
the experimental measurements of
the slope of the scalar form factor in $K_{\mu3}^L$ decay. 
In order to establish this discrepancy more precisely, we need an accurate direct measurement 
of lnC\footnote{This could allow a
matching with the two loops computation of $K_{l3}$ and the first model independent extraction
of 
$V^{us}$.},
avoiding the ambiguity attached to the slope measurement based on the 
use of theoretically flawed assumptions. 
If this disagreement with the SM is confirmed, it could be interpreted
in the framework of the LEET as a manifestation of physics beyond the SM by direct
right-handed couplings of fermions to W.
These new couplings, if they exist, have to appear in every process involving 
charged currents\footnote{For a complete 
analysis of CC interactions see~\cite{BOPS06}.}. 
Howe-ver they are not easy to disentangle from the extraction of the fundamental observables of 
QCD at low energy (form factors, $\alpha_S$, quark masses...).
There are not many processes beyond the one discussed in this talk
in which the possible enhancement of $\epsilon_S$ could be tested.
It is not the case for the hadronic $\tau$ decays, the $\nu$ ($\bar{\nu}$) DIS off valence quarks
or $K^0-\bar{K^0}$ mixing. Constraints arising from the CP violation sector could be interesting to study
in this framework.   
\\\textbf{Acknowledgements~:} I would like to thank the organizers for this enriching conference.

\end{document}